\documentclass[aps,pra,reprint,longbibliography,nofootinbib,showkeys]{revtex4-2}
\usepackage[T1]{fontenc}
\usepackage{amsmath,amssymb}
\usepackage{booktabs}
\usepackage{graphicx}
\usepackage{listings}
\usepackage{hyperref}
\hypersetup{colorlinks=true,linkcolor=blue,citecolor=blue,urlcolor=blue,
 pdftitle={Reducing Trust Cost in Quantum Application Development Using AI},
 pdfauthor={Katsuhiro Endo}}
\newcommand{\ket}[1]{\lvert #1\rangle}
\newcommand{\bra}[1]{\langle #1\rvert}
\newcommand{\norm}[1]{\lVert #1\rVert}
\newcommand{\enc}{\operatorname{enc}}
\newcommand{\code}[1]{\texttt{#1}}
\begin{document}
\title{Reducing Trust Cost in Quantum Application Development Using AI}
\author{Katsuhiro Endo}
\affiliation{National Institute of Advanced Industrial Science and Technology (AIST)}
\begin{abstract}
Generative AI can be a powerful tool for supporting the design and implementation of quantum applications, whereas the burden of checking whether generated circuits realize the intended computation increases. In this study, we regard this burden of human understanding, checking, and judgment as ``trust cost,'' and propose a development methodology that integrates AI and formal verification to reduce it. In the proposed development methodology, the implementation is divided into three components: a target mathematical specification, a quantum circuit, and a correspondence theorem and its proof that prove the agreement between the two. This division separates the specifications and theorem statements checked by humans from the theorem proofs entrusted to mechanical checking. As a concrete development example of this proposal, we describe an implementation for the time evolution of a stochastic particle system and its results.
\end{abstract}
\maketitle

\begin{figure*}[t]
\centering
\includegraphics[width=\textwidth]{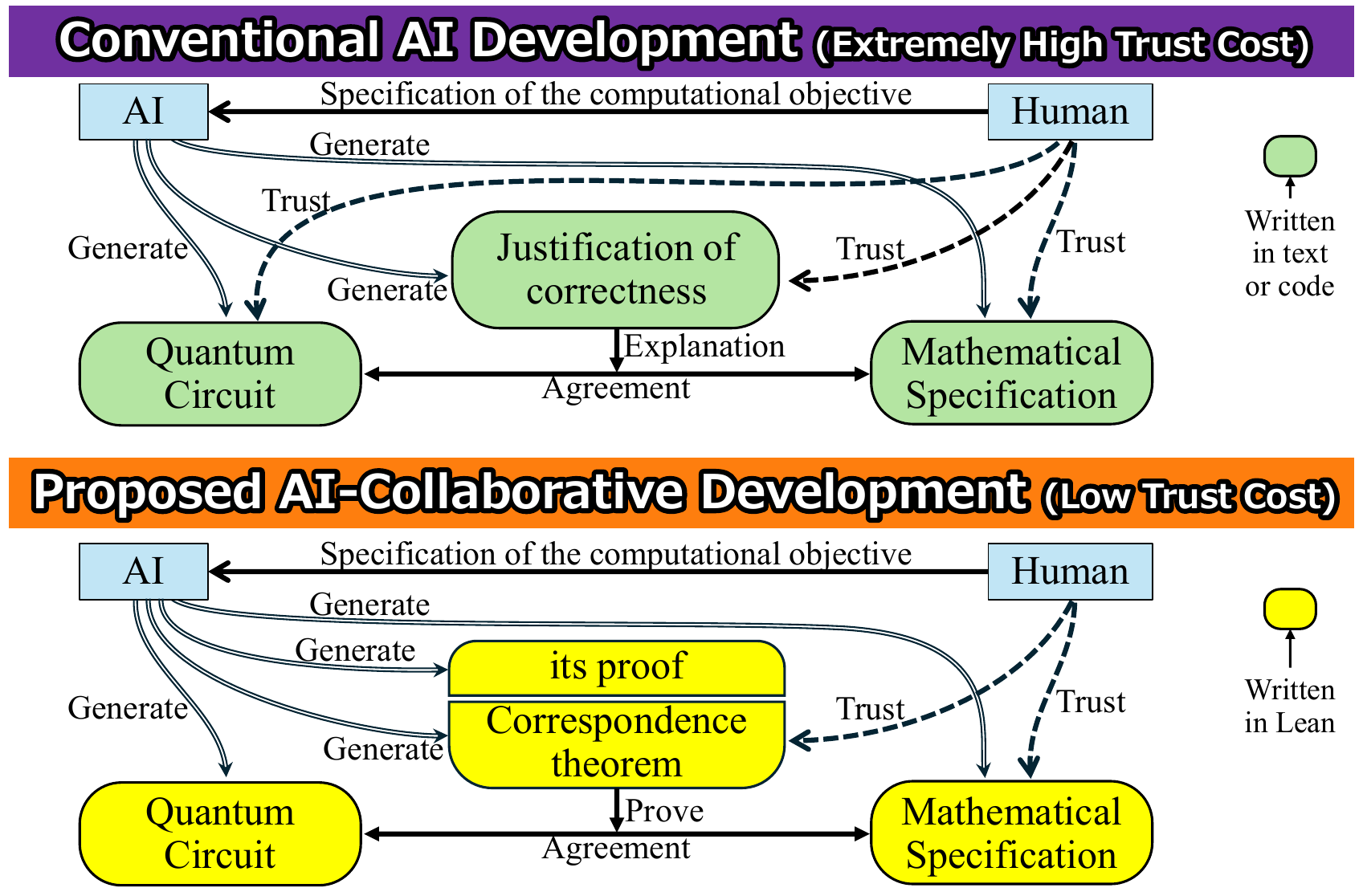}
\caption{Comparison of quantum application development methodologies using AI. Upper panel: conventional development using AI. The correctness of AI-generated artifacts cannot be confirmed unless all of them are correctly understood and trusted. Lower panel: the proposed AI-collaborative development methodology. Formal verification separates the parts that the developer must trust from those that can be verified automatically, thereby reducing trust cost.}
\label{fig:development-method}
\end{figure*}

\section{Introduction}
To clarify the usefulness of quantum computers, it is necessary to turn scientific and engineering problems into concrete quantum applications based on appropriate quantum algorithms. Quantum computer users expect quantum algorithms to reduce the computation time required to obtain their desired results compared with conventional classical algorithms. This means that it is necessary not only to show a speedup in asymptotic computational complexity (order) compared with classical algorithms, but also to numerically evaluate the computational resources required for a concrete problem size and required accuracy. In implementing quantum applications, constant factors omitted in asymptotic complexity and the burden of error correction, among other factors, change the conditions under which a theoretical speedup leads to a practical advantage~\cite{beverland,quetschlich}.

Therefore, quantum application developers need to translate each process of the algorithms constituting the quantum application into quantum circuits, so that their implementation costs (qubit count, gate count, and quantum circuit depth) can be examined. In fact, libraries for describing quantum algorithms and estimating their resources, such as Qualtran~\cite{harrigan}, are being developed. Needless to say, this development work requires broad expertise in quantum computing, including quantum algorithms and quantum circuit implementation. On the other hand, quantum application developers should have expertise in the applications they wish to develop, such as quantum chemistry or mechanical engineering. Requiring additional expertise in quantum computing raises the barrier to quantum application development itself. To further accelerate quantum application development and efficiently accelerate the development of use cases for quantum computers, it is important to lower this development barrier.

Meanwhile, the recent progress of AI technology has been remarkable, and expectations are growing that AI support for design and implementation can also be applied to quantum application development. In fact, even within the author's observations, examples of generating quantum circuits that execute a desired computation and designing which quantum algorithms are needed have rapidly increased over the past year. As an example of prior research, the QCoder Benchmark~\cite{mikuriya} proposes quantum code generation from tasks described in natural language and code correction based on execution results from a simulator.

However, research and development artifacts generated by AI are not necessarily correct. Even if an AI-generated circuit is executable, whether the quantum circuit realizes the intended computation is a separate issue and needs to be checked. In addition, checking concrete examples for which the correct answers are known in advance (test cases) does not guarantee that the computation is performed correctly for all input cases.

Thus, even if using AI for quantum application development reduces the burden of algorithm development and programming, it creates a need to scrutinize the AI's claims in detail in order to trust the correctness of its artifacts. This ultimately requires expertise to understand the correspondence between the behavior of the quantum circuit and the intended computation. Therefore, even if AI simply replaces human development work, the barrier of requiring quantum circuit expertise from application developers remains if checking the results depends on human understanding and trust.

In this study, we refer to this burden of human understanding, checking, and judgment as ``trust cost.'' In other words, to use AI in quantum application development, research and development of engineering methodologies that coexist with AI are needed, addressing how to introduce a development methodology that lowers this ``trust cost.''

Accordingly, in this study, we propose a development methodology using formal verification by theorem proving as a way to reduce the ``trust cost'' of quantum application development using AI. Formal verification is a method that mathematically describes the behavior of a program or another object and the properties it should satisfy, and verifies their correspondence based on logic. Previous applications of formal verification to quantum programs include Qbricks~\cite{chareton}, which addresses the verification of parameterized circuit-building programs, and SQIR~\cite{hietala}, which verifies algorithm correctness using a proof assistant. In this study, we propose a quantum application development methodology premised on formal verification and generative AI for reducing ``trust cost.'' By defining the division of roles among the parts trusted by humans, the parts developed by AI, and the parts guaranteed by theorem proving, expertise in the detailed implementation of quantum circuits becomes unnecessary. We also apply this methodology to the development of quantum circuits that simulate classical stochastic diffusion phenomena on a quantum computer, and present the results.

\section{Proposal of a Development Methodology for Reducing Trust Cost}
For a formally stated theorem, whether a formally given proof correctly proves the theorem can be checked mechanically. On the other hand, whether the theorem statement appropriately represents the developer's objective requires trust based on the developer's understanding. If a theorem different from the developer's intention is formally stated, providing its proof does not mean that the developer's intended objective has been achieved.

The same applies to theorem and proof descriptions produced with AI assistance. Therefore, to judge whether a theorem written by AI appropriately represents the developer's intention, the developer must check the theorem statement, including the definitions referred to by the theorem, and this cannot be avoided. Of course, AI can generate an explanation of a theorem statement to aid understanding, but it is the developer's responsibility to trust that the statement fits the objective. Preparing another AI for auditing and checking the implementation may help discover differences between the developer's intention and the theorem statement. However, without direct checking by the developer, this merely amounts to unconditionally trusting the auditing AI, and does not constitute verification of the correctness of the AI-generated artifact.

In contrast, having AI write the proof of a theorem is essentially different from having AI write the theorem itself. A proof is a description that guides formal verification software in mechanically deriving the correctness of the theorem. Thus, regardless of how the proof is written, the theorem is necessarily correct as long as the formal verification software can mechanically judge it to be correct. (Of course, this does not hold if there is a bug in the implementation of the formal verification software; here, we trust the formal verification software adopted in this study and assume that it contains no errors.) This means that the developer does not need to know at all how the proof of the theorem is written. Thus, even if a proof is given by AI (even if it is an extremely unusual proof), the developer does not need to trust the proof at all. The correctness of the theorem is established without requiring trust in the proof.

Therefore, reducing the trust cost of quantum application development by combining AI and theorem proving requires defining development requirements that specify which parts the developer understands and trusts, and which parts require no trust at all.

In the development requirements proposed in this study, we first divide the development items into the following three components. The first is the ``mathematical specification,'' which describes the computational objective formally and in a form understandable to the developer. The second is the ``quantum circuit,'' which formally describes the quantum circuit. The third is the ``correspondence theorem and its proof,'' which states the correspondence between the ``mathematical specification'' and the ``quantum circuit.''

Figure~\ref{fig:development-method} compares conventional AI-assisted quantum application development with the proposed methodology.

The first component, the ``mathematical specification,'' is a description that defines the action, state, distribution, or other object to be realized by the quantum algorithm and quantum circuit independently of the quantum circuit. The ``mathematical specification'' prioritizes what the developer wants to realize and is written in a form that is clear and easy for the developer to understand. We require this description to contain no expertise concerning the concrete construction of quantum circuits and to be understandable based on domain knowledge of the target application. The mathematical specification must also not refer to either of the other two components, the ``quantum circuit'' and the ``correspondence theorem and its proof.''

The second component, the ``quantum circuit,'' is defined as a gate sequence describing on which qubits and in which order the most basic gates (hereafter, primitive gates) are executed, or as a function that generates such a gate sequence. The quantum circuit provides a concrete quantum circuit implementation for realizing the mathematical specification. The developer does not need to understand at all how this quantum circuit is concretely given, but only needs to trust that it is a sequence composed solely of primitive gates. The quantum circuit must also not refer to either of the other two components, the ``mathematical specification'' and the ``correspondence theorem and its proof.''

The third component, the ``correspondence theorem and its proof,'' consists of a formal theorem statement describing how the mathematical specification and the quantum circuit correspond, and a description proving that this correspondence holds. The correspondence theorem refers to the target mathematical specification and quantum circuit and shows their agreement under the conditions specified by the theorem. This correspondence describes, for example, agreement between the quantum circuit regarded as an action and the specification defined by the mathematical specification, or agreement between the state after executing the quantum circuit and the mathematical specification. Agreement here may be approximate, with a specified required error. The developer must understand and trust that this theorem statement represents the intended correspondence. In contrast, human trust in the proof body that establishes the correspondence is unnecessary; checking by formal verification software serves this role.

When describing the correspondence theorem, the developer does not need to derive the operator or state generated by the quantum circuit by tracing the individual gates in the referenced circuit. The developer only needs to understand and trust the common definitions that convert a circuit description into a mathematical action. These definitions consist only of (i) the definition of the operator corresponding to each primitive gate constituting the quantum circuit, and (ii) the definition that, when two different quantum circuits are combined, the operator corresponding to the resulting circuit is the product of the operators corresponding to the original circuits. By recursively applying (ii), the operator corresponding to any quantum circuit composed of primitive gates is decomposed into a product of operators corresponding to primitive gates.

Mathematically, a correspondence theorem formally defines the following:
\begin{equation}
 \forall\theta,\quad V(\theta)\Longrightarrow
 R\bigl([\![C(\theta)]\!],S(\theta)\bigr).
 \label{eq:correspondence}
\end{equation}
Here, $\theta$ denotes the input parameters, $V(\theta)$ the conditions that $\theta$ should satisfy, $C(\theta)$ a function that generates a quantum circuit when input parameters are given, $[\![\,\cdot\,]\!]$ a function that converts a quantum circuit into an operator, $S(\theta)$ the mathematical specification, and $R$ the relation (such as agreement) that the circuit action and the mathematical specification should satisfy. The input parameters may include the input conditions of the target system and the required error to be achieved by the circuit.

As briefly described above, the scope of the developer's understanding for these three components can be summarized as follows. For a mathematical specification, the developer understands and trusts its entire definition, recursively including the definitions it refers to. For a quantum circuit, the developer only checks and trusts the fact that the description constructs a circuit from primitive gates. Its concrete contents do not need to be trusted. Specifically, as described below, the form of the quantum circuit is guaranteed through type checking in a programming language for formal verification. For a correspondence theorem, the developer understands and trusts the entire theorem statement, recursively including the definitions it refers to. In contrast, for the proof of the theorem, its contents do not need to be trusted as long as the theorem has been formally verified.

Dividing the development items into these three components, and defining the mathematical specification and the quantum circuit independently of the other two components, is extremely important for reducing trust cost. If the mathematical specification is defined by referring to part of the quantum circuit, the developer ultimately needs to understand and trust the circuit definition as well in order to fully trust the mathematical specification. By having the mathematical specification and the quantum circuit referred to together for the first time in the correspondence theorem statement, the cost of trusting this quantum circuit can be eliminated entirely.

\section{Example Target for Quantum Application Development}
This section describes the application to be developed, in order to report a concrete quantum application example using the development methodology proposed in the preceding section.

\subsection{Overview of the Target Computation on a Quantum Computer}
In this study, we consider the following stochastic time-evolution system. At most $K$ particles exist on a $G\times G$ square lattice, where $G=2^q$. Periodic boundary conditions are imposed, and lattice sites at the upper, lower, left, and right boundaries are adjacent to those at the opposite boundaries. We assume that each lattice site is occupied by at most one particle. Each particle moves with probability $\Delta t$ during an infinitesimal time interval $\Delta t$, randomly moving one lattice spacing either to the right or upward, with probability $1/2$ for each direction. If a particle exists at the destination, the moving particle and the particle at the destination coalesce into one particle. Each coalescence decreases the particle count by one.

\begin{figure}[t]
\centering
\includegraphics[width=\columnwidth]{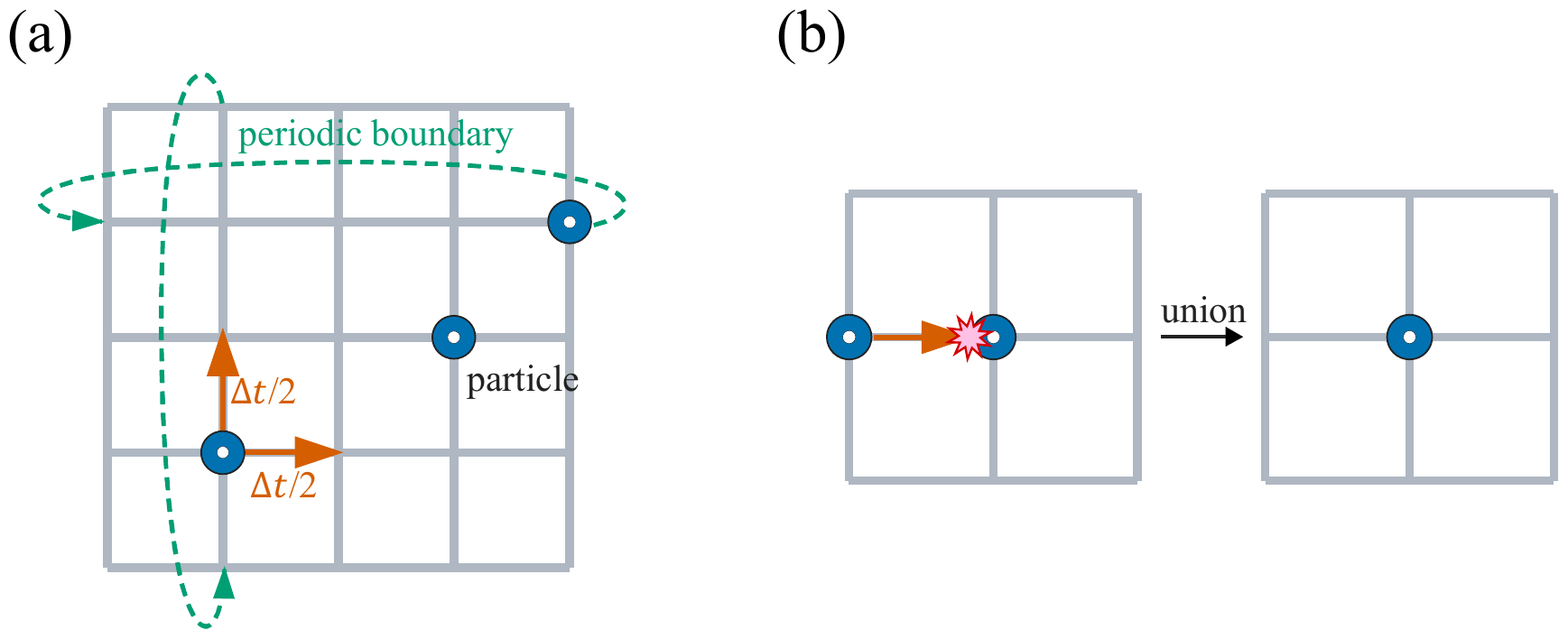}
\caption{Definition of the time evolution of the classical stochastic particle diffusion system used in the quantum application development example. (a) During an infinitesimal time interval $\Delta t$, each particle moves one lattice spacing upward or to the right with total probability $\Delta t$. The two directions are chosen with equal probability. (b) If another particle occupies the destination, the two particles coalesce into one.}
\label{fig:particle-dynamics}
\end{figure}

The particle motion and coalescence rules are illustrated in Fig.~\ref{fig:particle-dynamics}.

Let $\Omega$ be the set of all possible particle configurations on the lattice. If $P(s;t)$ is the probability of configuration $s\in\Omega$ at time $t$, the time evolution of the state of this system is described by a differential equation in time for $P(s;t)$. Namely, for any configurations $s,v\in\Omega$, with the generator matrix of the system denoted by $Q$, we have
\begin{equation}
 \frac{d}{dt}P(v;t)=\sum_s Q_{v,s}P(s;t).
 \label{eq:master}
\end{equation}
From the behavior of the system, $Q_{v,s}$ is derived as
\begin{equation}
 Q_{v,s}=
 \begin{cases}
 -|s| & v=s,\\
 \frac{1}{2}\operatorname{Count}(v,s) & v\ne s.
 \end{cases}
 \label{eq:generator}
\end{equation}
Here, $|s|$ is the number of particles in configuration $s$, and $\operatorname{Count}(v,s)$ is the total number of moves that reach configuration $v$ when one particle in configuration $s$ is selected and moved upward or to the right (the number of patterns yielding $v$ among the particle count times two possible moves). The diagonal elements of the generator matrix represent probability outflow from that configuration, and the total transition rate out of the configuration increases in proportion to the number of particles. The off-diagonal elements represent inflow into other configurations, with probabilities proportional to the number of paths. Thus, if the differential equation for time evolution in Eq.~(\ref{eq:master}) can be solved, the state distribution at an arbitrary time can be calculated. In this example, we consider a quantum computation that obtains the configuration probability distribution $P(s;T)$ at a specified time $T$ from a given initial state distribution $P(s;0)$.

\subsection{Quantum Algorithm Used}
This subsection describes the quantum algorithm used in the present quantum application development example to calculate the time evolution of the probability distribution defined in the preceding subsection. This quantum algorithm is a partial excerpt from a method proposed by the author in a separate paper~\cite{endo}.

First, to represent the probability distribution in Eq.~(\ref{eq:master}) on a quantum computer, we encode each configuration of the system as a fixed-length bit string. If we were to store, for each lattice site $i$, a one-bit state $n_i\in\{0,1\}$ indicating whether a particle is present, representing a configuration would consume $G^2=4^q$ qubits, and the required qubit count would increase exponentially with $q$. Therefore, using the fact that there are at most $K$ particles, we use the following $(2qK+1)$-qubit encoding:
\begin{equation}
 \ket{\enc(s)}
 =\ket{w_1}\ket{w_2}\cdots\ket{w_r}\ket{1}
   \ket{0}^{\otimes 2q(K-r)}.
 \label{eq:encoding}
\end{equation}
Here, $w_j$ is the binary representation of the coordinates $(x_i,y_i)$ of the $j$th particle occupying a lattice site in configuration $s$. If $x_i$ and $y_i$ each range from $0$ to $G-1$, $w_j$ can be represented by $2\log G=2q$ bits. Also, $r=|s|\le K$ is the number of particles in configuration $s$. This encoding $\enc(s)$ places a $1$ immediately after the $r$ coordinates $w_1w_2\cdots w_r$, each of $2q$ bits, and fills the remaining $2q(K-r)$ bits with $0$, giving a total of $2qK+1$ qubits. This encoding is useful when the maximum number of particles is sufficiently small.

In this encoding, the number of particles can be identified from the number of consecutive zeros at the end, so any two different configurations are guaranteed to have different encodings. Since the $1$ acts as a delimiter, we hereafter call this encoding the delimiter-based configuration encoding. Furthermore, to guarantee the uniqueness of the mapping in the delimiter-based configuration encoding, the $r$ coordinates $w_1w_2\cdots w_r$ are arranged in lexicographic coordinate order. This guarantees that any configuration has one particular encoding.

Since the delimiter-based configuration encoding gives a bit representation of a particular configuration, the probability distribution of configurations representing the state of the system can be encoded in amplitudes as
\begin{equation}
 \ket{p(t)}=\sum_{s\in\Omega}P(s;t)\ket{\enc(s)}.
 \label{eq:distribution-vector}
\end{equation}
It is clear from the definition that $\ket{p(t)}$ is an $N=2^{2qK+1}$-dimensional vector. Note that its norm is at most one. With the encoding of configurations, the system generator matrix $Q$ can also be represented as a matrix on the encoded space. We define the encoded generator matrix $\widetilde Q\in\mathbb R^{N\times N}$ by
\begin{equation}
 \widetilde Q_{a,b}=
 \begin{cases}
 Q_{v,s} & \enc(v)=a\ \land\ \enc(s)=b,\\
 0 & \text{otherwise}.
 \end{cases}
 \label{eq:encoded-generator}
\end{equation}
The time evolution of $\ket{p(t)}$ is then written as
\begin{equation}
 \frac{d}{dt}\ket{p(t)}=\widetilde Q\ket{p(t)}.
 \label{eq:encoded-evolution}
\end{equation}
This can be solved analytically as
\begin{equation}
 \ket{p(t)}=e^{\widetilde Q t}\ket{p(0)}.
 \label{eq:analytic-evolution}
\end{equation}

For the computations described below, the quantum algorithm used here implements a quantum circuit whose operator is a unitary matrix $U_Q$ that block-encodes $\widetilde Q$. Namely, the matrix $U_Q$ satisfies
\begin{equation}
 \bigl(\bra{0^{\otimes n_{\mathrm{anc},Q}}}\otimes I\bigr)
 U_Q
 \bigl(\ket{0^{\otimes n_{\mathrm{anc},Q}}}\otimes I\bigr)
 =\frac{\widetilde Q}{\alpha_Q}.
 \label{eq:generator-block}
\end{equation}
Here, $n_{\mathrm{anc},Q}$ is the number of ancilla qubits required for the block encoding, and $\alpha_Q$ is the scaling factor when embedding $\widetilde Q$ in $U_Q$. These are constants determined for a concrete quantum circuit.

As a time-evolution method using this block-encoding implementation, we introduce the linear matrix equation for solving time evolution proposed by Jennings et al.~\cite{jennings}, and solve it using the linear system solver proposed by Dalzell~\cite{dalzell}. In the method of Jennings et al., time evolution is calculated by solving the linear matrix equation
\begin{equation}
 Hx=b.
 \label{eq:history-system}
\end{equation}
Giving the definitions directly, $x$ and $b$ are defined as
\begin{equation}
 \begin{aligned}
 x&=(X_0,X_1,\ldots,X_{c-1})^{\mathsf T},\\
 X_j&=(x_{j,0},x_{j,1},\ldots,x_{j,d})^{\mathsf T},
 \end{aligned}
 \label{eq:history-vector}
\end{equation}
and
\begin{equation}
 b=(\ket{p(0)},0,\ldots,0)^{\mathsf T}.
 \label{eq:rhs}
\end{equation}
Here, $c\ge1$ and $d\ge0$ are integers; $\ket{p(0)}$ and $x_{j,k}$ are $N$-dimensional vectors, $X_j$ is a $(d+1)N$-dimensional vector, and $x,b$ are $c(d+1)N$-dimensional vectors. The matrix $H$ is the $c(d+1)N\times c(d+1)N$ matrix defined by
\begin{equation}
 H=
 \begin{pmatrix}
 D_0&O&\cdots&\cdots&O\\
 -F_0&D_1&\ddots&&\vdots\\
 O&-F_1&\ddots&\ddots&\vdots\\
 \vdots&\ddots&\ddots&D_{c-2}&O\\
 O&\cdots&O&-F_{c-2}&D_{c-1}
 \end{pmatrix}.
 \label{eq:history-matrix}
\end{equation}
Here, $D_i$ and $F_i$ are defined as
\begin{equation}
 D_i=
 \begin{cases}
 \begin{pmatrix}
 I_N&O&\cdots&\cdots&O\\
 -\frac{h\widetilde Q}{1}&I_N&\ddots&&\vdots\\
 O&-\frac{h\widetilde Q}{2}&\ddots&\ddots&\vdots\\
 \vdots&\ddots&\ddots&I_N&O\\
 O&\cdots&O&-\frac{h\widetilde Q}{d}&I_N
 \end{pmatrix} & i<m,\\[6pt]
 \begin{pmatrix}
 I_N&O&\cdots&\cdots&O\\
 -I_N&I_N&\ddots&&\vdots\\
 O&\ddots&\ddots&\ddots&\vdots\\
 \vdots&\ddots&-I_N&I_N&0\\
 O&\cdots&O&-I_N&I_N
 \end{pmatrix} & i\ge m,
 \end{cases}
 \label{eq:diagonal-block}
\end{equation}
and
\begin{equation}
 F_i=
 \begin{cases}
 \begin{pmatrix}
 I_N&I_N&\cdots&I_N\\
 O&O&\cdots&O\\
 \vdots&\vdots&\ddots&\vdots\\
 O&O&\cdots&O
 \end{pmatrix} & i<m,\\[6pt]
 \begin{pmatrix}
 O&\cdots&O&I_N\\
 O&\cdots&O&O\\
 \vdots&\ddots&\vdots&\vdots\\
 O&\cdots&O&O
 \end{pmatrix} & i\ge m.
 \end{cases}
 \label{eq:forward-block}
\end{equation}
Here, $I_N$ is the $N\times N$ identity matrix, $O$ is the zero matrix, $h$ is a positive real constant, and $m$ is an integer constant satisfying $0\le m<c$.

The linear matrix equation defined above satisfies, for $i<m$,
\begin{equation}
 x_{i+1,0}=\left(\sum_{k=0}^{d}\frac{(h\widetilde Q)^k}{k!}\right)x_{i,0}.
 \label{eq:taylor-step}
\end{equation}
This is the degree-$d$ Taylor approximation of $e^{\widetilde Qh}$. Thus, the vector $x_{i,0}$ is the simulation result after $i$ steps of the degree-$d$ explicit Taylor method with time step $h$ and initial state $x_{0,0}=\ket{p(0)}$. For $i\ge m$, the relation
\begin{equation}
 \begin{gathered}
 x_{i,j}=x_{i',j'}\\
 \text{for }m\le i,i'<c\text{ and }0\le j,j'\le d
 \end{gathered}
 \label{eq:idling}
\end{equation}
holds. This means that the time-evolution result $x_{m,0}$ up to step $m$ is copied exactly to the subsequent variables. This copying is known to be useful when extracting the state at the final time.

Next, we wish to obtain the solution $x$ of the linear matrix equation defined in Eq.~(\ref{eq:history-system}) as a quantum state. To do this, Dalzell's linear system solver considers an extension of the state space by one additional dimension:
\begin{equation}
 H^{\mathrm{ext}}=
 \begin{pmatrix}H/\alpha_H&0\\0&\nu^{-1}\end{pmatrix},
 \quad
 \ket{b^{\mathrm{ext}}}=\frac{1}{\sqrt2}
 \begin{pmatrix}b/\norm b\\1\end{pmatrix}.
 \label{eq:extension}
\end{equation}
Here, $\alpha_H$ is the scaling constant for the block encoding of $H$, and $\nu$ is an arbitrary positive constant. Based on this extended matrix and vector, consider the matrix
\begin{equation}
 B=\bigl(I-\ket{b^{\mathrm{ext}}}\bra{b^{\mathrm{ext}}}\bigr)H^{\mathrm{ext}}.
 \label{eq:b-matrix}
\end{equation}
For the vector
\begin{equation}
 z=(H^{\mathrm{ext}})^{-1}\ket{b^{\mathrm{ext}}}
   =\frac{1}{\sqrt2}\begin{pmatrix}u\\\nu\end{pmatrix},
 \quad u=\frac{\alpha_H}{\norm b}H^{-1}b,
 \label{eq:kernel-vector}
\end{equation}
it can be shown that
\begin{equation}
 \ker B=\operatorname{span}\{z\}.
 \label{eq:kernel}
\end{equation}
This means that the right singular vector corresponding to the zero singular value of $B$ is proportional to $z$.

From this relation, if we first prepare a block encoding of $B$ and transform all singular values by the function
\begin{equation}
 f(x)=
 \begin{cases}
 1 & x=0,\\
 -1 & x\ne0,
 \end{cases}
 \label{eq:reflection-function}
\end{equation}
using quantum singular value transformation (QSVT)~\cite{gilyen}, we can realize a block encoding of
\begin{equation}
 R=2\ket z\bra z-I.
 \label{eq:reflection}
\end{equation}
Here, $\ket z=z/\norm z$. Equation~(\ref{eq:reflection}) shows that $R$ is a kernel reflection action that reverses the $z$ component of a state vector. Thus, applying $R$ from the right to the unit state vector whose only nonzero component is a $1$ in the extended part,
\begin{equation}
 \ket{e^{\mathrm{ext}}}=(0,\ldots,0,1)^{\mathsf T},
 \label{eq:extended-basis}
\end{equation}
gives
\begin{equation}
 \begin{aligned}
 R\ket{e^{\mathrm{ext}}}
 &=\bigl(2\ket z\bra z-I\bigr)\ket{e^{\mathrm{ext}}}\\
 &=\frac{1}{\norm u^2+\nu^2}
   \begin{pmatrix}2\nu u\\\nu^2-\norm u^2\end{pmatrix}.
 \end{aligned}
 \label{eq:reflected-state}
\end{equation}
Since $u$ is precisely a quantity proportional to the desired solution ($x=H^{-1}b$), in particular, when $\nu=\norm u$, the normalized solution state of $x$ can be obtained with norm one (success probability one). In an actual quantum circuit implementation, applying $f(x)$ by QSVT requires a polynomial approximation, and hence the implementation is approximate, based on the required accuracy.

In practical quantum computation, further quantum computations are performed on the solution state $x$, and quantities useful for the quantum application are measured. Since the purpose of this study is to give a concrete example of the proposed development methodology, our implementation target is limited to the quantum circuit that obtains this solution state $x$.

\section{Development Example Based on the Proposed Methodology}
This section presents a development example for the target in the preceding section, in the form proposed in this study. The mathematical specifications, quantum circuits, and correspondence theorems and their proofs are written in Lean~\cite{lean}, a programming language for formal verification. Descriptions that humans understand and trust are separated from descriptions for which mechanical checking by Lean alone is sufficient.

\subsection{Separating the Lean Implementation into Three Components and Implementing Mathematical Operators in Lean}
Developing the quantum application described in the preceding section requires implementing the generator matrix $\widetilde Q$, the matrix $H$ of the linear equation, the kernel reflection matrix $R$, and the final solution state $x$ using quantum circuits. In the proposed development methodology, we first define these as checkpoints with independent mathematical specifications. For each checkpoint, we construct the mathematical specification serving as the implementation target, the quantum circuit that achieves it, and a theorem stating whether the circuit correctly realizes the specification. This allows us to check whether the correct specification is realized for each meaningful implementation unit. In addition, by not making only the final objective a mathematical specification, we can prevent the unit of specification to be understood at one time from becoming too large for the developer. The checkpoints in this development are designed as follows:
\begin{itemize}
 \item CP01: encoded generator matrix $\widetilde Q$;
 \item CP02: matrix $H$ of the linear equation;
 \item CP03: kernel reflection matrix $R$;
 \item CP04: final solution state $x$.
\end{itemize}

The developer first describes a mathematical specification in Lean for each of these checkpoints. AI may perform this writing itself. However, the developer needs to check whether the specification written by AI is correct. Below, we explain each Lean implementation mainly for CP01.

The top-level description of the mathematical specification of CP01 was written in Lean as the function \code{generatorMatrix}, as follows (partially simplified for readability):
\par\noindent\begin{minipage}{\linewidth}
\begin{lstlisting}
def generatorMatrix (p : CWM.Parameters) :
    Matrix (DelimiterWord p) (DelimiterWord p) ℂ :=
  fun out inp =>
    match decodeConf p out, decodeConf p inp with
    | some outConf, some inConf =>
        (CWM.generator outConf inConf : ℂ)
    | _, _ => 0
\end{lstlisting}
\end{minipage}\par
Omitting the details, this states that the encoded $Q$ matrix associates an entry with the corresponding element of $Q$ only when the configuration encodings can be correctly decoded to their original configurations. The final line states that a matrix entry is zero if decoding is not successful. The function \code{CWM.generator} outputs an element of $Q$, and \code{decodeConf} restores a configuration from its encoding. Naturally, \code{CWM.generator} and \code{decodeConf} are also parts of the mathematical specification that the developer must understand. For example, \code{CWM.generator} is defined as
\par\noindent\begin{minipage}{\linewidth}
\begin{lstlisting}
def CWM.generator {p : CWM.Parameters} :
    Matrix (Configuration p) (Configuration p) ℝ :=
  fun out inp =>
    if out = inp then -(inp.1.card : ℝ)
    else (transitionCount out inp : ℝ) / 2
\end{lstlisting}
\end{minipage}\par
Here, \code{inp.1.card} is the number of particles in the configuration, and \code{transitionCount} is further defined through nested definitions as the number of path patterns reaching configuration \code{out} from configuration \code{inp}. This is exactly the definition of the $Q$ matrix described in Sec.~III. Namely, the mathematical specification of CP01 is the procedure for constructing the $Q$ matrix itself. In this way, mathematical specifications use intuitive but rigorous definitions that are easy for developers to understand.

\subsection{Implementation of Quantum Circuits in Lean}
On the other hand, the quantum circuit of CP01 was defined in Lean as follows:
\par\noindent\begin{minipage}{\linewidth}
\begin{lstlisting}
def generatorCircuit (p : GeneratorParameters) :
    Circuit( PrimitiveGate( (generatorQubits p) ) ) :=
  [output omitted]
\end{lstlisting}
\end{minipage}\par
This Lean definition receives input parameters and outputs a quantum circuit. Here, how the circuit is implemented is not important to the developer; what matters is only whether it consists solely of a gate sequence composed of primitive gates. If this output contains elements unrelated to quantum circuits or gates other than primitive gates, the developer must check whether it is a correctly operating quantum circuit. Conversely, as long as it is guaranteed to be an executable quantum circuit, whether it correctly realizes the target mathematical specification can be checked through the ``correspondence theorem and its proof.''

Thus, in the proposed development methodology, automatic checking by types is used to guarantee that an executable quantum circuit is output. In the definition of \code{generatorCircuit} above, the output type is specified as \code{Circuit(PrimitiveGate(...))}. Therefore, the developer can trust this by simply checking that the definition of this type accepts only sequences of primitive gates.

In fact, \code{Circuit} and \code{PrimitiveGate} are independently defined as follows:
\par\noindent\begin{minipage}{\linewidth}
\begin{lstlisting}
inductive PrimitiveGate (qubits : Nat) where
  | x (target : Fin qubits)
  | h (target : Fin qubits)
  | ry (angle : AngleExpr) (target : Fin qubits)
  | phaseShift (angle : AngleExpr) (target : Fin qubits)
  | cnot [arguments omitted]
  | toffoli [arguments omitted]
inductive Circuit (Gate : Type*) where
  | identity
  | gate (value : Gate)
  | seq (first second : Circuit Gate)
  | repeat (count : Nat) (body : Circuit Gate)
\end{lstlisting}
\end{minipage}\par
First, \code{PrimitiveGate} represents the list of gates allowed within quantum circuits. The present definition shows that the type can take only the six primitive gates listed. Other gates are not included as primitive gates. The \code{Circuit} type takes an arbitrary \code{Gate} type as an argument and is defined to take one of the following four forms: (i) identity (\code{identity}), (ii) one element of type \code{Gate} (\code{gate}), (iii) an ordered sequence of two elements of type \code{Circuit} (\code{seq}), or (iv) repetition of one element of type \code{Circuit} a specified number of times (\code{repeat}). Thus, \code{Circuit} is a syntax tree constructed recursively using circuits of type \code{Circuit}, with gates of type \code{Gate} at its leaves.

By checking these definitions, the developer can trust that \code{Circuit(PrimitiveGate(...))} is a quantum circuit consisting only of gates of type \code{PrimitiveGate}. Note, however, that these definitions alone do not describe the action by which the circuit changes a quantum state.

How the action of a quantum circuit is defined is likewise something the developer should trust, but it is independent of the internal construction of individual quantum circuits. This transformation into an action is defined by the following \code{denote} function:
\par\noindent\begin{minipage}{\linewidth}
\begin{lstlisting}
def denote (gateOp : Gate → Operator basis) :
    Circuit Gate → Operator basis
  | .identity => 1
  | .gate value => gateOp value
  | .seq fi se => denote gateOp se * denote gateOp fi
  | .repeat count body => (denote gateOp body) ^ count
\end{lstlisting}
\end{minipage}\par
The \code{denote} function receives ``a function that takes a value of type \code{Gate} and returns an action (\code{Operator})'' and generates a function that outputs an action from a value of type \code{Circuit(Gate)}. The important point here is that the action of a quantum circuit (\code{Circuit(Gate)}) is determined indirectly from ``a function that takes a value of type \code{Gate} and returns an action (\code{Operator}),'' rather than directly from the quantum circuit itself. With this construction, the developer can indirectly trust the action of \code{Circuit} simply by understanding and trusting the actions of primitive gates and the rules for composing actions in the \code{denote} function.

In the present implementation, a function called \code{primitiveGateOperator} is defined as the ``function that takes a value of type \code{Gate} and returns an action (\code{Operator}).'' The actions of the $X$ gate, $H$ gate, and so on are individually defined as matrix actions on state vectors. These functions, \code{denote} and \code{primitiveGateOperator}, are used to state agreement theorems in the ``correspondence theorem and its proof.''

\subsection{Implementation of Correspondence Theorems and Their Proofs in Lean}
The ``correspondence theorem and its proof'' of CP01 is constructed by referring to the mathematical specification \code{generatorMatrix} and the quantum circuit \code{generatorCircuit} of CP01 described above:
\par\noindent\begin{minipage}{\linewidth}
\begin{lstlisting}
theorem generatorCircuit_correct [arguments omitted]:
    BlockReadout.squareBlock
      (generatorBinding p)
      (denote primitiveGateOperator
        (generatorCircuit [arguments omitted]))
    =	((3 * K))⁻¹ • generatorMatrix p := by
  [proof body omitted]
\end{lstlisting}
\end{minipage}\par
This theorem shows that the encoded block of the operator of the quantum circuit generated by \code{generatorCircuit} equals $1/(3K)$ times the mathematical specification defined by \code{generatorMatrix}. Briefly, \code{BlockReadout.squareBlock} is a generic function that extracts the part of an action corresponding to specified input and output bases. Here, it is used to extract the part where the ancillas are zero, $(\bra{0_{\mathrm{anc}}}\otimes I)U(\ket{0_{\mathrm{anc}}}\otimes I)$, where $\mathrm{anc}$ denotes the bits corresponding to ancillas and $U$ is the target action. Namely, \code{generatorCircuit} generates a circuit of type \code{Circuit(PrimitiveGate(...))}, which \code{denote} converts into an action using \code{primitiveGateOperator}. Then, \code{BlockReadout.squareBlock} extracts the encoded block from the resulting circuit action. This corresponds to the left-hand side of the equality ($=$) in the theorem. The right-hand side is a constant multiple of the action given by \code{generatorMatrix} as the mathematical specification. Thus, equality between the two sides guarantees that the action realized by the circuit agrees without error with the mathematically defined specification. In addition, the right-hand side gives the specific constant factor $1/(3K)$, where $K$ is the maximum number of particles. This gives a concrete scaling constant for the block encoding implemented by CP01. Therefore, the present implementation proves that the concrete quantum circuit realizes a block encoding with scaling constant $\alpha_Q=3K$. This can be regarded as a benefit of implementing even the quantum circuit implementation in Lean.

The theorem description shown above is the content of the theorem itself; its proof is omitted here. Lean can mechanically check whether this theorem is correct based on a given proof. Regardless of whether AI wrote the proof body, the developer can trust the correctness of this theorem without reading the contents of the proof. Therefore, in the ``correspondence theorem and its proof,'' it is sufficient to check that the theorem correctly reflects the developer's intention. Of course, this additionally requires checking that the subordinate elements, including \code{BlockReadout.squareBlock}, are correctly defined.

\subsection{Common Foundation of the Proposed Development Methodology and a Proposed Method for Introducing QSVT}
The preceding discussion shows that what developers should understand and trust includes not only Lean descriptions specific to individual applications but also Lean descriptions that can be used in common across multiple application developments. Examples are \code{PrimitiveGate}, \code{Circuit}, \code{denote}, and \code{BlockReadout.squareBlock}. Since the same definitions can be used to describe mathematical specifications and correspondence theorems even when the computational target changes, these also help reduce the trust cost of each application when the proposed methodology is used for multiple application developments in the future. Note, however, that moving a definition to the common foundation does not remove it from the scope of the developer's understanding and trust. After checking the meaning of a description defined as part of the common foundation, the developer can reuse that understanding when placing trust in uses of the same definition.

As one part of this common foundation, we propose how to handle QSVT, a general-purpose quantum algorithm. The application targeted in this development also uses QSVT to construct the kernel reflection action $R$. To actually construct a quantum circuit using QSVT, the following processes are needed in order: (a) derivation of the polynomial to apply, (b) numerical calculation of the phases for quantum signal processing (QSP) that realize this polynomial, and (c) construction of the quantum circuit from the numerical phase sequence obtained. If the objective is simply to execute a concrete quantum circuit, QSP phases can be obtained by numerical calculation using dedicated software~\cite{dong}.

In contrast, in a formal correspondence proof, simply substituting a numerically obtained phase sequence into the circuit does not guarantee that the circuit satisfies the required error. Proving a correspondence theorem for a quantum circuit based on a numerical phase sequence requires evaluating the effect of numerical errors in the phases on the circuit action, making the theorem proof difficult. However, defining the phase sequence itself analytically also creates the difficulty that circuit construction and theorem proving become too complicated.

Therefore, in this proposal, we separately define the existence of a QSP phase sequence that realizes the required polynomial and the actual numerical calculation of a phase sequence. In the formal verification of a theorem, if the existence of a value satisfying specified conditions has been proved, subsequent claims can be proved using those conditions without calculating the concrete value. For example, when $a^2=9$, this corresponds to obtaining the value of $a^4$ without calculating the concrete value of $a$, as long as the existence of a solution can be shown.

We therefore introduce the \code{defaultPhaseProvider} function, which takes a required polynomial as input and allows a phase sequence realizing that polynomial transformation to be handled without obtaining concrete numerical values. Based on a proof that at least one QSP phase sequence realizing the required polynomial exists, this function outputs a value representing one phase sequence satisfying the conditions. Although this value is not given as concrete numerical values, a proof exists that this phase sequence realizes the required polynomial, so it can be used in subsequent proofs.

In the Lean functions that generate the quantum circuits of CP03 and CP04, circuits can be defined using the value returned by \code{defaultPhaseProvider}, and the correspondence between those circuits and their mathematical specifications can be proved. This is because the action of the entire quantum circuit can be described mathematically without obtaining the concrete numerical values of the phase sequence.

On the other hand, independently of a correspondence theorem, one may wish to input a concrete phase sequence and generate a concrete quantum circuit. In this case, another function that outputs a concrete numerical sequence can be supplied as an argument to the circuit-generating function in place of \code{defaultPhaseProvider}. Even when the way the phase sequence is provided is replaced, the primitive gate count and qubit count of the generated circuit remain unchanged if the phase sequence has the same length. Of course, in this case, the proof of the correspondence theorem is no longer valid, but providing a phase sequence that realizes the required polynomial with sufficient accuracy becomes the responsibility of the numerical software that calculates the phase sequence.

For example, the quantum circuit definition of CP03 is as follows:
\par\noindent\begin{minipage}{\linewidth}
\begin{lstlisting}
def kernelReflectionCircuit {AngleType : Type}
    (input : FiniteCircuitInput)
    (phaseProvider : FinitePhaseProvider AngleType) :
    Circuit (PrimitiveGate input.kernelQubits) :=
  [concrete circuit definition omitted]
\end{lstlisting}
\end{minipage}\par
Here, \code{FinitePhaseProvider} is the type of a function that receives a request for a polynomial to be achieved and returns a phase sequence. It is an abstract type that allows either a function producing a numerically obtained phase sequence or \code{defaultPhaseProvider} to be passed to the same circuit-construction function.

\subsection{Implementation of All Checkpoints Based on the Proposed Development Methodology Using AI}
The preceding subsections described how the proposed development methodology should be implemented using Lean. This subsection describes how the present development example was implemented. In this development, generative AI was used for all Lean implementations of all mathematical specifications, quantum circuits, and correspondence theorems and their proofs from CP01 to CP04. The author considered the computational objective and method in advance, and requested all Lean implementations after providing generative AI with a description of the implementation target roughly as described in Sec.~III and a description of the proposed development methodology roughly as described in Sec.~II.

However, having AI implement the code in Lean alone may not result in mathematical specifications and correspondence theorems implemented as the author intended. The author therefore checked whether all Lean implementations of the mathematical specifications and correspondence theorems were as intended. As discussed above, the developer does not need to inspect the internal implementations of the quantum circuits or the proof bodies of the correspondence theorems. Since a single implementation request did not completely achieve what is described in Sec.~II, revisions were repeated so that the mathematical specifications and correspondence theorems would be described as intended and defined in a way understandable to humans. This development was conducted by the author alone, and took approximately one month. Most of this month consisted of repeated work by generative AI to revise the proof implementations until they compiled in Lean. (Successful compilation means that the correctness of the theorem has been checked, provided, of course, that no unintended axioms have been introduced and no unproved assumptions using \code{sorry} have been used.)

During development, we observed cases in which generative AI concealed discovered problems inside mathematical specifications and correspondence theorems in order to superficially fulfill the author's instructions. This can be regarded as an example showing that the developer must check and trust all of the mathematical specifications and correspondence theorems.

The Lean source code, including the mathematical specifications,
quantum circuits, correspondence theorems and their proofs,
and resource-count verification, is available at
\url{https://github.com/Katsuhiro-Endo/qlean-trust-cost}.

\section{Applications and Limitations of the Proposed Development Methodology}
\subsection{Exact Resource Estimation Using the Lean Implementation}
In this development methodology, the implementation for realizing the target mathematical specification is obtained as a Lean implementation of a quantum circuit composed of primitive gates. Therefore, using this circuit definition, gate counts and qubit counts for concrete inputs can be evaluated exactly.

We present the results of evaluating the quantum gate count and qubit count required for the circuit implementation of the present development target under specific input conditions. Here, the quantum gate count is the number of primitive gates contained in the circuit, without distinguishing gate types. Evaluation using other definitions is also possible. The qubit count is the total including ancilla qubits used for auxiliary calculations. Note that it is not necessary to actually generate the gate sequence when evaluating quantum gate counts or similar quantities. Separately from each checkpoint's theorem, a gate-count formula can be constructed in Lean and a theorem that it equals the gate count of the checkpoint's quantum circuit can be proved, allowing the exact gate-count formula to be trusted. In this study, AI constructed this theoretical formula and its theorem proof, and the author understood and trusted the Lean statement of the theorem.

As a concrete example of input parameters for the evaluated computation, we set $q=2$ ($G=2^q=4$), $K=2$, $m=1$, $h=1/6$ ($T=1/6$), $d=1$, and $c=3$. We set the required error $\epsilon$ of the kernel reflection action to $10^{-2}$, where the spectral-norm bound between the true reflection action $R$ and the implemented reflection action $R'$ is $\norm{R-R'}_2\le\epsilon$. The initial state of the system consists of two particles, one at the upper-leftmost lattice site and the other at the lower-rightmost lattice site.

\begin{table}[t]
\caption{Evaluation of quantum resources for concrete parameters.}
\label{tab:resources}
\begin{ruledtabular}
\begin{tabular}{lrr}
Quantity constructed & Gate count & Qubit count\\
\hline
$Q$ (CP01) & 16,041 & 124\\
$H$ (CP02) & 40,197 & 275\\
$R$ (CP03,04) & 417,312,971,982,202 & 3,342\\
\end{tabular}
\end{ruledtabular}
\end{table}
The evaluation results for the quantum circuits are shown in Table~\ref{tab:resources}. The quantities constructed are block encodings of $Q$, $H$, and $R$, respectively, and the table shows the gate counts and qubit counts for implementing them. The table shows that implementing the kernel reflection action $R$ requires an extremely large gate count and qubit count.

The reason for the increase in gate count is that the QSVT degree required for the theorem achieving the required error is set to $3{,}125{,}001$. CP03 proves that this degree achieves the required error. However, whether a degree of approximately three million is actually necessary is a separate issue. The current theoretical formula proposed by AI takes the required degree to be the smallest odd integer greater than $2(d+4)^{c(d+1)}/\epsilon$. Since this increases exponentially with $c$ and super-exponentially with $d$, it may be an extremely loose upper bound. A more rigorous theoretical estimate may allow a polynomial degree bound to be achieved.

Regarding the qubit count, while only 12 system qubits are required to represent the solution $x$, the number of qubits required to construct $R$ reaches approximately 280 times that number. We found that the main reason for this increase is that the instructions given to generative AI did not impose a limit on the number of auxiliary qubits. As a result, the implementation requires separate auxiliary qubits instead of reusing qubits that can be reused, for example by reusing qubits set to zero by uncomputation at later stages. Appropriate implementation revisions may also lead to a substantial reduction in the qubit count.

\subsection{Limitations of the Proposed Development Methodology}
As described in the preceding subsection, the quantum circuits generated by generative AI have room for further improvement. Even if the target mathematical specification is correctly defined, freedom remains in selecting theoretical parameters and circuit constructions needed for an actual quantum circuit implementation. Therefore, further research and development are needed to search for better implementation methods in terms of quantum resources.

However, it is also conceivable that AI could automate the search for theories and implementations aimed at reducing quantum resources. Since the present development methodology is proposed from the viewpoint of reducing trust cost, it would be appropriate to proceed toward more practical implementations by using this methodology as a basis while simultaneously using AI to search for reductions in quantum resources.

\section{Conclusion}
In this study, we regarded the burden on developers of understanding, checking, and judging AI-generated artifacts in quantum application development using AI as ``trust cost,'' and proposed a development methodology to reduce it. While AI is an excellent technology with the potential to open quantum application development to everyone, its inappropriate use may instead increase the final burden on developers. Therefore, the advancement of engineering methodologies that appropriately use AI is an urgent task. We hope that this proposal will contribute to quantum application development that combines the use of AI's generative capabilities with a reduction in the burden on developers.

\begin{acknowledgments}
This work was partially supported by JST [Moonshot R\&D] [Grant Number JPMJMS256J].

For the preparation of this English manuscript, AI was used solely to translate the original Japanese manuscript into English.
\end{acknowledgments}

\bibliographystyle{apsrev4-2}
\nocite{*}
\bibliography{qlean-trust-cost-references}
\end{document}